\documentclass[twocolumn,showpacs,preprintnumbers,amsmath,amssymb]{revtex4-1}

\usepackage{graphicx}
\usepackage{dcolumn}
\usepackage{bm}
\usepackage{appendix}
\usepackage{chngcntr}
\usepackage{multirow}
\usepackage{diagbox}
\usepackage{color}

\begin{document}
	
	\title{Optical information encryption using general temporal ghost imaging with practical experimental condition}
	\author{Juan Wu${}^{1,2}$}
	\author{Xiaohai Zhan${}^{2,3}$}
	\author{Fang-Xiang Wang${}^{2,3}$}
	\email{fxwung@ustc.edu.cn}
	\author{Zhenqiu Zhong${}^{2,3}$}
	\author{Shuang Wang${}^{2,3}$}
	\author{Wei Chen${}^{2,3}$}
	\author{Zhen-Qiang Yin${}^{2,3}$}
	\author{Zheng-Fu Han${}^{2,3}$}
	\affiliation{${}^1$College of Electrical and Information Engineering, Anhui University of Science and Technology, Huainan 232001, China}
	\affiliation{${}^2$CAS Key Laboratory of Quantum Information, University of Science and Technology of China, Hefei 230026, China}
	\affiliation{${}^3$CAS Center for Excellence in Quantum Information and Quantum Physics, University of Science and Technology of China, Hefei 230026, China}

	
	\begin{abstract}
Temporal Ghost Imaging (TGI), which reconstructs fast temporal signals using a slow detector, holds significant potential in optical communication, high-speed imaging, and quantum information processing. However, achieving high-quality information reconstruction has been a major challenge for the practical application of TGI. A theoretical model [\emph{ Applied Optics}, 62(5): 1175-1182 (2023)] was proposed to investigate the influence of experimental parameters of the slow detector on image quality; however, its experimental verification was hitherto lacking. In this study, we implemented a multi-bit information transmission scheme based on both quantum and classical TGI methods. Experimental validation confirmed the accuracy of the theoretical model and demonstrated its application in encrypting noisy multi-bit information. The experimental results demonstrate that as the information coding density increases, the decoding accuracy becomes highly sensitive to the detection accuracy and threshold of the slow detector. When these parameters degrade, the decoding quality deteriorates significantly. Additionally, our system shows notable robustness against detection noise, but loses the ability to accurately decode when the noise amplitude becomes too high. Our work endows TGI with optical information encryption capabilities in practical systems and furnishes comprehensive guidelines for the further application of TGI.
		
	\end{abstract}
	
	\pacs{Valid PACS appear here}
	\maketitle
	
	
	\section{\label{sec:level1}Introduction}
 In today's rapidly advancing technological landscape, the continuous evolution of communication techniques has driven significant progress in optical information encryption, a critical component of information security. The growing demand for secure data transmission has prompted the exploration and adoption of various encryption methods \cite{SC1,SC2,SC3,SC4,SC5}. However, the performance of optical encryption and transmission systems remains highly susceptible to channel disturbances, which can compromise the integrity and confidentiality of the encrypted data, ultimately undermining the overall security of the system \cite{SC2,SC5}. This challenge can be addressed through the introduction of the Ghost Imaging (GI) scheme \cite{secure1,secure2,secure3}. A key feature of GI is that the reconstruction of the target object depends on the detection results from both the reference path and test path, forming the foundation of its secure encryption and transmission capabilities \cite{spacedomain1,spacedomain2,spacedomain3,timedomain1,timedomain2,timedomain3}. Over the past decade, the concept of optical encryption via GI has been extensively studied, evolving from the space domain \cite{GIcode1,GIcode2,GIcode3,GIcode4,GIcode5,GIcode6} to the time domain \cite{TSC1,TSC2,TSC3,TSC4,TSC5,TSC6}. Temporal Ghost Imaging (TGI) has garnered considerable attention due to its exceptional noise robustness and potential applications in fields such as quantum secure communication \cite{SC5,secure1,DeviceEvaluation} and multi-wavelength imaging \cite{TSC3,TSC4,wavelength1,wavelength2,wavelength3}.

Image quality is the most critical factor in evaluating the efficacy and performance of a Temporal Ghost Imaging (TGI) system. The pursuit of high-quality imaging within a reduced sampling time has driven the development and innovation of various enhancement techniques. To improve imaging quality, several methods have been proposed and demonstrated, including signal-to-noise ratio (SNR) enhancement schemes \cite{MaginifiedTGI,DTGI,FTGI}, parallel measurement technologies \cite{SpaceMultiplexing,WavelengthMultiplexing,FrequencyMultiplexing}, and one-time readout temporal single-pixel imaging \cite{OTRTGI1,OTRTGI2}. However, a major challenge in TGI systems arises from the random intensity fluctuations of the light source. These fluctuations introduce redundant information, which reduces the correlation efficiency between the reference and object arms, ultimately degrading image quality. To address this issue, we have developed a theoretical model to thoroughly analyze the impact of intensity accuracy and the detection threshold of the slow detector on TGI image quality \cite{model}. Nevertheless, experimental validation of the slow detector's impact on optical image encryption and transmission has been lacking, creating a significant research gap that this study aims to fill.

In this study, we have implemented the quantum TGI (QTGI) and classical TGI (CTGI) multi-bit information transmission system to ensure the proper generation, transmission, and reception of signals. The slow detector, which is a critical component in generating the cipher text, is characterized and calibrated to quantify its impact on the overall system performance. The detection accuracy and detection threshold of the slow detector are precisely controlled and adjusted, and the effects of the decoding are quantified by the decoding accuracy rate (DAR). The experimental results demonstrate that the DAR is sensitive to detection accuracy and threshold of the slow detector with increasing information coding density. A significant improvement in DAR is observed when high-precision detectors are used near the binary detection threshold. The differential method is then implemented in both the CTGI and QTGI systems, and the resulting changes in DAR and transmission quality are carefully measured and analyzed. The experimental results  show that the differential method has a marginal effect on DAR in the CTGI system and no optimization effect in the QTGI system. To evaluate the anti-noise ability of the encryption system, controlled amounts of optical and electronic noise are introduced into the experimental setup. The system's performance is evaluated by observing the degradation in the reconstructed signals and the corresponding impact on DAR. The experimental results confirm the strong anti-noise ability of the TGI system as long as the target signal is not overwhelmed by noise. Our work provides an important guiding model and key experimental techniques for the practical development of optical information encryption transmission.

\section{Experimental demonstration of general theoretical model}

\begin{figure*}[ht]
	\centering
	\includegraphics[width=0.98\textwidth]{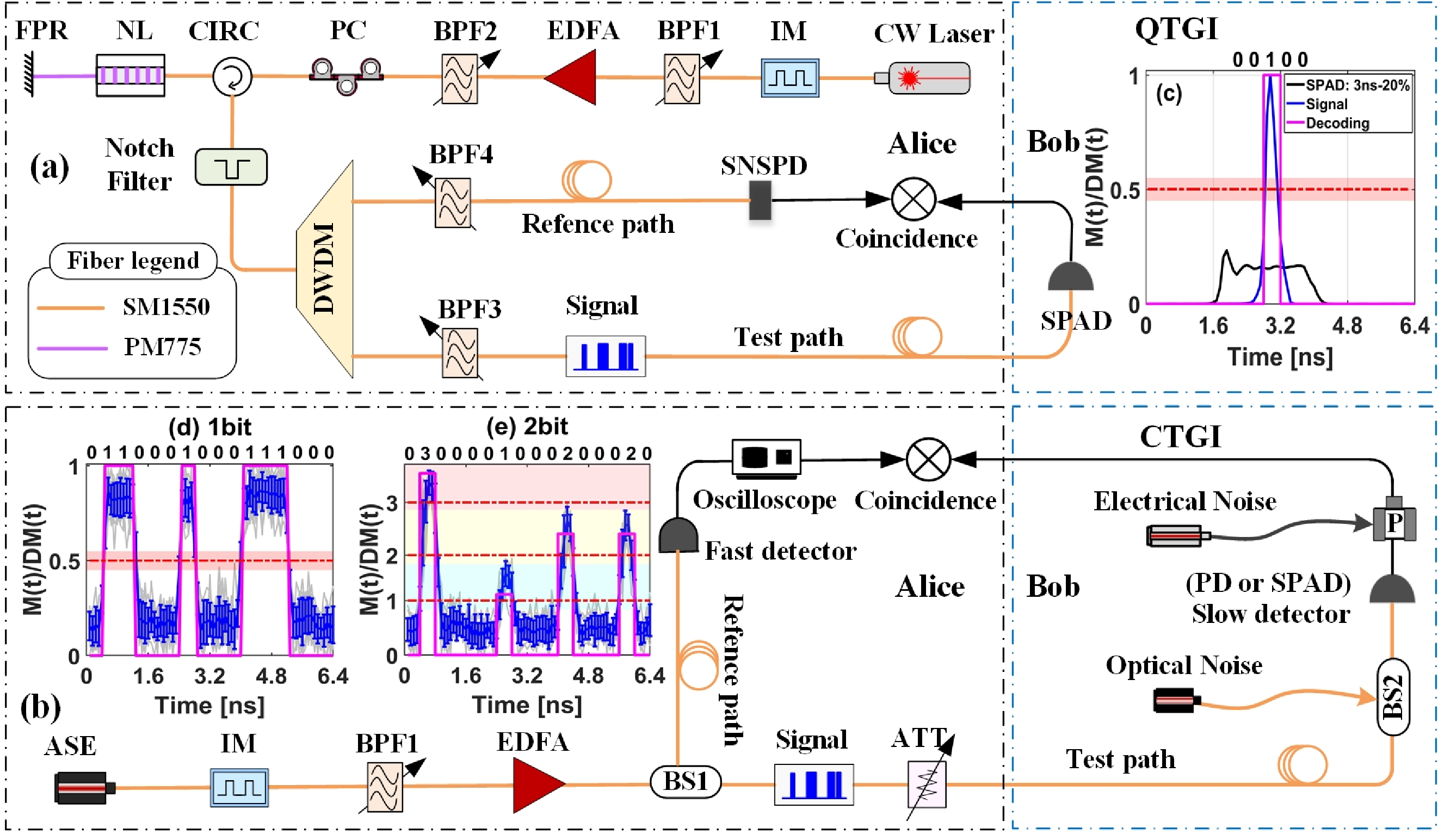}
	\caption{ QTGI (a) and CTGI (b) signal reconstruction and encryption transmission system. The nonlinear module(NL) comprises a fiber-pigtailed periodically poled lithium niobate waveguide (ppLN) and a temperature control module for regulating the second harmonic generation (SHG) conversion efficiencies. (c) Temporal structure of target signal detected by a high-speed detector (blue line) and its corresponding decoding results (magenta line). The black line is the detection efficiency of SPAD, where 3ns-20$\%$ means a gate width of 3 ns and a detection efficiency of 20$\%$. (d) and (e) represent the temporal structure of 1-bit and 2-bit signals, respectively. Gray background, reconstructed signals of 20 measurements; blue line, mean values of these 20 measurements; dotted red line, dividing line for decoding; magenta line, the decoding results.   }
	\label{fig-1} 
\end{figure*}

The QTGI signal reconstruction and encryption transmission system is depicted in Fig.\ref{fig-3} (a), utilizes a telecom CW pump that is modulated into pulses by an intensity modulator (IM) to ensure compatibility with the coherence time of the light source. The pulsed source operates at a repetition rate of 10 MHz and is then amplified by an erbium-doped fiber amplifier (EDFA). To achieve a high-quality source, background noise is filtered out using two band-pass filters (BPF). The photon pair source is created through a two-stage cascaded parametric conversion process. Initially, the pump pulse is directed into a fiber-pigtailed periodically poled lithium niobate (ppLN) waveguide (Covesion MSHG1550 with 40 mm long type-0 crystal), a commercially available second harmonic generation (SHG) module. The resulting 775 nm pulses are then reflected by the fiber pigtailed retroreflector (FPR, Thorlabs P5-780PMR-P01-1) to serve as the pump for the Type-0 SPDC process. The down-converted photon pairs are rerouted via the circulator (CIRC) followed by a notch filter to remove any remaining fraction of the pump or additional noise. Finally, the frequency-correlated photons are separated by a dense wavelength-division multiplexer (DWDM) into signal and idler photons. The idler photons are incident into the reference path as the secret key $I{_r}(t)$ and detected by a superconducting nanowire single-photon detector (SNSPD) with high resolution. The signal photons are entered into the target signal as ciphertext $I{_t}(t)$ and subsequently detected a gated-mode single-photon avalanche detector (SPAD, Qasky, WT-SPD300-LN) without temporal resolution in the timing windows. The detection signals of both the SNSPD and SPAD are synchronously recorded by a time-to-digital converter (TDC). The repetition rate of the detection system is 10 MHz. Additional bandpass filters (BPF3 and BPF4) are added to further suppress the background noises before the detectors. It is important to note that both the IM and the PPLN waveguide module operate in the slow axis only, necessitating the addition of a polarization controller (PC) after the IM to minimize the insertion loss. After N times of synchronous measurements, then the temporal signals can be reconstructed according to the correlation function of the TGI \cite{DeviceEvaluation,model}, which is defined by
\begin{equation}
	M(t) \propto {\rm{cov}}({I_r}{I_t}) = {\left\langle {{I_{coin}}} \right\rangle _N}
	\label{eq:TGI}
\end{equation}
where ${\left\langle {} \right\rangle _N}$ is the ensemble average over N times of synchronous measurements. As the SAPD executes click detection, the imaging system does not have any fluctuations in light intensity \cite{model}. Therefore, the target signals can be reconstructed by counting the coincidence ($I_{coin}$) of signal and idler photons.

The CTGI signal reconstruction and encryption transmission system is shown in Fig. \ref{fig-1}(b). The transmitter(Alice) prepares a temporally randomized source and keeps the reference path of the joint TGI. The amplified spontaneous emission (ASE) is first being chopped into pulses by an intensity modulator (IM), and subsequently filtered by a 100 GHz bandwidth dense wavelength division multiplexer (DWDM). The repetition rate of the pulsed source is 10 MHz. An erbium-doped fiber amplifier (EDFA) is used to amplify the output intensity. The pulsed source is then divided into the reference path and the test path by a 99:1 unpolarized beam splitter (BS1). The light from the reference path is detected by a fast detector and an real-time oscilloscope to record the source pattern distribution with the bandwidth of 5 GHz and 1.25 GHz, respectively. The source field is considered as the secret key $I{_r}(t)$ that is shared via the private channel. The test path's light will be entered into the encoding object, forming the ciphertext, which is subsequently transmitted to the receiver(Bob). The total intensity of the ciphertext ${I_t} = \int {{I_r}(t)M(t)dt} $ is detected by a slow detector (SPAD or 35MHz PD). Then, Bob sends the detected results of the PDs to Alice through the public channel. The Att is a tunable attenuator utilized for regulating the optical power entering the detector. After N times of synchronous measurements, Alice and Bob reconstruct the target object according to the joint and local monitoring results. The normalized joint correlation function can be expressed as
\begin{equation}
	M{(t)_{TGI}} = \frac{{{\mathop{\rm cov}} {{({I_r}{I_t})}}}}{{\sqrt {D({I_r}) \times D({I_t})} }}
	\label{EQ1}
\end{equation}
where $D(I_x) = \left\langle {{{[\Delta {I_x}]}^2}} \right\rangle$, $x = r,t$. $I_{r}$ and $I_{t}$ are the measured intensities of the fast and slow photon detectors, respectively. ${\rm{cov}}({I_r}{I_t}) = \left\langle {\Delta {I_r}\Delta {I_t}} \right\rangle  = \left\langle {{I_r}{I_t}} \right\rangle  - \left\langle {{I_r}} \right\rangle \left\langle {{I_t}} \right\rangle $, $\Delta I_r(t) = I_r(t) - {\left\langle {I_r(t)} \right\rangle}$ and $\Delta I_t=I_t-\langle I_t\rangle$ are measured intensity fluctuations. 

In order to meet the high efficiency and the low noise requirements in encryption system, SPAD generally avalanche in specific timing windows. In this mode, the temporal detection efficiency of the gated-mode SPAD can be regarded as a temporal object, and then the detailed process of the SPAD inside the gating windows can be reconstructed. Therefore, to mitigate the influence of SPAD's detection efficiency on information transmission and decoding, the temporal structure of the target signal $[M(t)]$ in the QTGI and CTGI system is designed to be more compact than that of the SPAD and is positioned within the stable region inside the gate windows of the SPAD, as illustrated by the magenta line in Figure 1(c). The nominal parameter of SPAD is set as a gate width of 3 ns and a detection efficiency of 20$\%$. The temporally detection efficiency of the SPAD is calibrated by scanning the delay time of an ultra-narrow pulsed laser \cite{c1}. The target signals are generated by an IM under the control of an arbitrary function generator (AFG) with a bandwidth of 240 MHz, yielding a minimum pulse width of approximately 800 ps for the target signal, as depicted by the blue line in Figure 3(c).

When the CTGI encryption system transmits multi-bit information, the 1-bit and 2-bit temporal signals are randomly encoded as '0110001000111000' and '0300001000200020', respectively, as shown in Fig. \ref{fig-1}(b) and (c). The reconstructed results of 20 measurements are represented by the gray background, while the mean values of these measurements are shown in the blue line. The effective characteristic time of intensity fluctuation is approximately 800 ps, which corresponds to the upper temporal resolution of the TGI system. To enhance the decoding process, a strategy of decoding every five data points from the reconstructed signals is adopted. All 20 sets of experimental signals were successfully decoded, as demonstrated by the magenta line, with each set showing agreement with their respective original signals. Randomly selected dividing lines within the partition zone are depicted by dotted red lines. For decoding 1-bit signals, it is determined whether the measured intensity exceeds or falls below these divider lines; if it exceeds, a decoding result of 1 is assigned; otherwise, it is assigned as 0. The same rule applies to decoding 2-bit signals.

\begin{figure*}[h]
	\centering
	\includegraphics[width=0.98\linewidth]{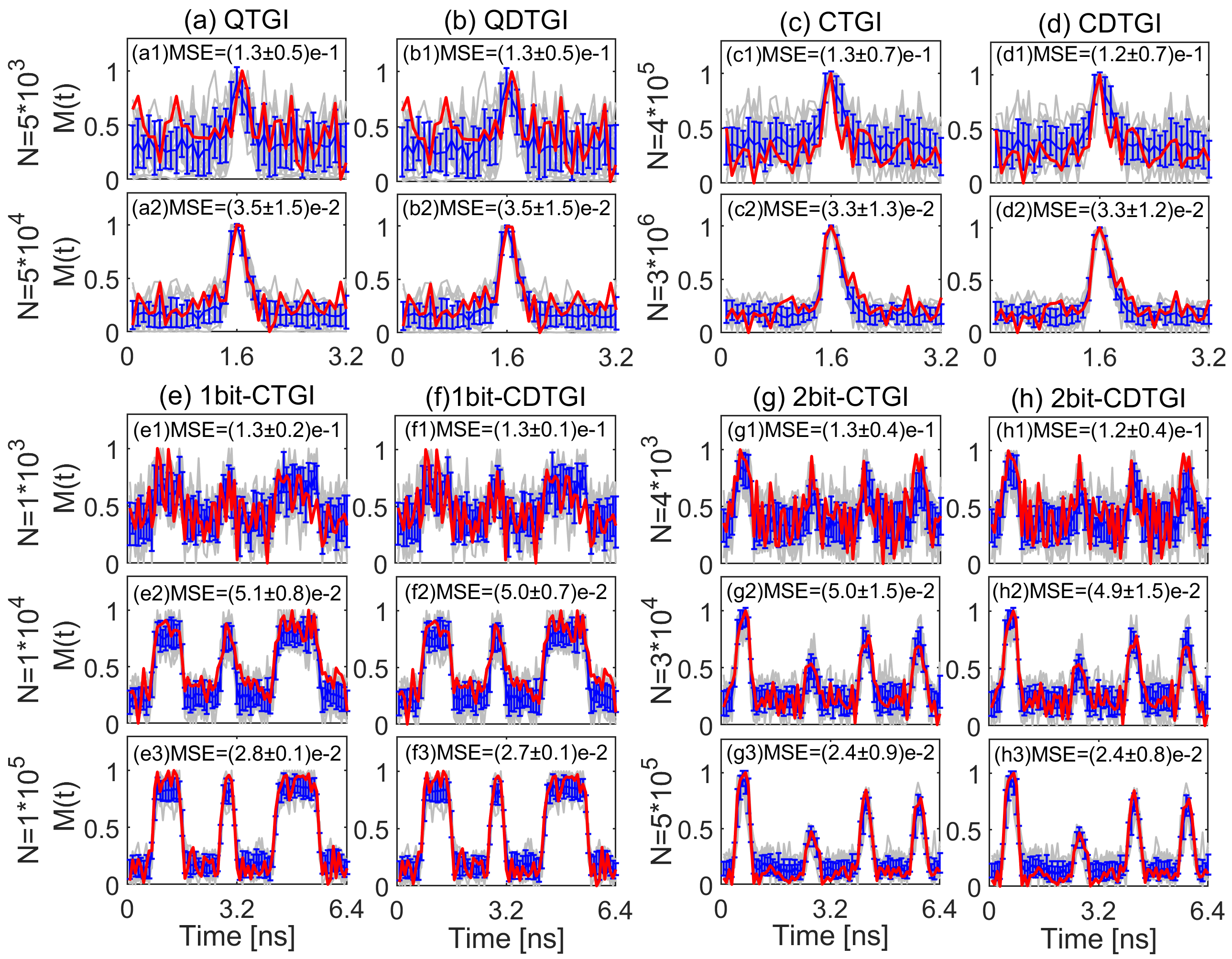}
	\caption{Reconstructed $M(t)$ of Q(D)TGI and C(D)TGI. Gray background, reconstructed signals of 20 measurements; blue line, mean values of the 20 measurements; red line, a typical signal acquisition. }
	\label{fig-2}
\end{figure*}

Figure \ref{fig-2} (a)-(d) present the reconstruction quality of $M(t)$ for C(D)TGI and Q(D)TGI. To attenuate the influence of the SPAD's temporal characteristics on the reconstruction process, a 1-bit '010' signal is employed as the target signal. It is evident that mean squared error (MSE) decreases as the number of synchronous measurements $N$ increases. For Q(D)TGI, the reconstruction images closely match the target signals when the number of realizations $N$ reaches $5\times10^4$, whereas C(D)TGI necessitates $N=3\times10^6$ measurements to attain an equivalent image quality. This is principally due to the quantum characteristics manifested by quantum correlated photon pairs, which can effectively alleviate background noise interference within the imaging system. These outcomes furnish compelling verification of the efficacy of the QTGI scheme in diminishing measurement requisites and data acquisition time. Additionally, it can be noted that, when considering an equivalent quantity of data, the reconstructed image of QTGI remains congruent with that of QDTGI, and the image quality of CDTGI is only marginally, rather than substantially, better than that of CTGI. The experimental results signify that DTGI does not exhibit distinct advantages over TGI.

To quantify the influence of the slow detector on the decoding quality, we further demonstrate the multi-bit encryption system using CTGI. Fig. \ref{fig-2} (e)-(f) give the multi-bit reconstructed images $[M(t)]$ with high detection accuracy of slow PD ($W=100$, where $W$ is the detection accuracy division of the slow detector). It is clearly observed that when $N$ reaches $10^4$, the MSE of the 20 reconstruction results for the 1-bit signal is $(5.1 \pm 0.8)\times{10^{-2}}$. At this moment, the reconstructed image exhibits a relatively low noise level and presents a satisfactory fitting performance with the target signal. However, to achieve a quality level comparable to that of a 1-bit signal, it is requisite to increase the measurement number of a 2-bit signal to $3\times10^4$. This is because that the random variations in signal amplitude necessitate an increment in the number of measurements in order to distinctly discriminate the disparities between the heights of the two signals. Therefore,the number of measurement $N$ reaches $5\times10^5$ for 2-bit signal, the reconstructed image can attain a fluctuation level comparable to that of 1-bit signals. Besides, we can also observe that the differential method shows a slight improvement in image quality. These experimental outcomes comprehensively and conclusively demonstrate that DTGI does not show clear advantages over TGI when transmitting multi-bit information. 

\begin{figure*}[ht]
	\centering
	\includegraphics[width=0.98\linewidth]{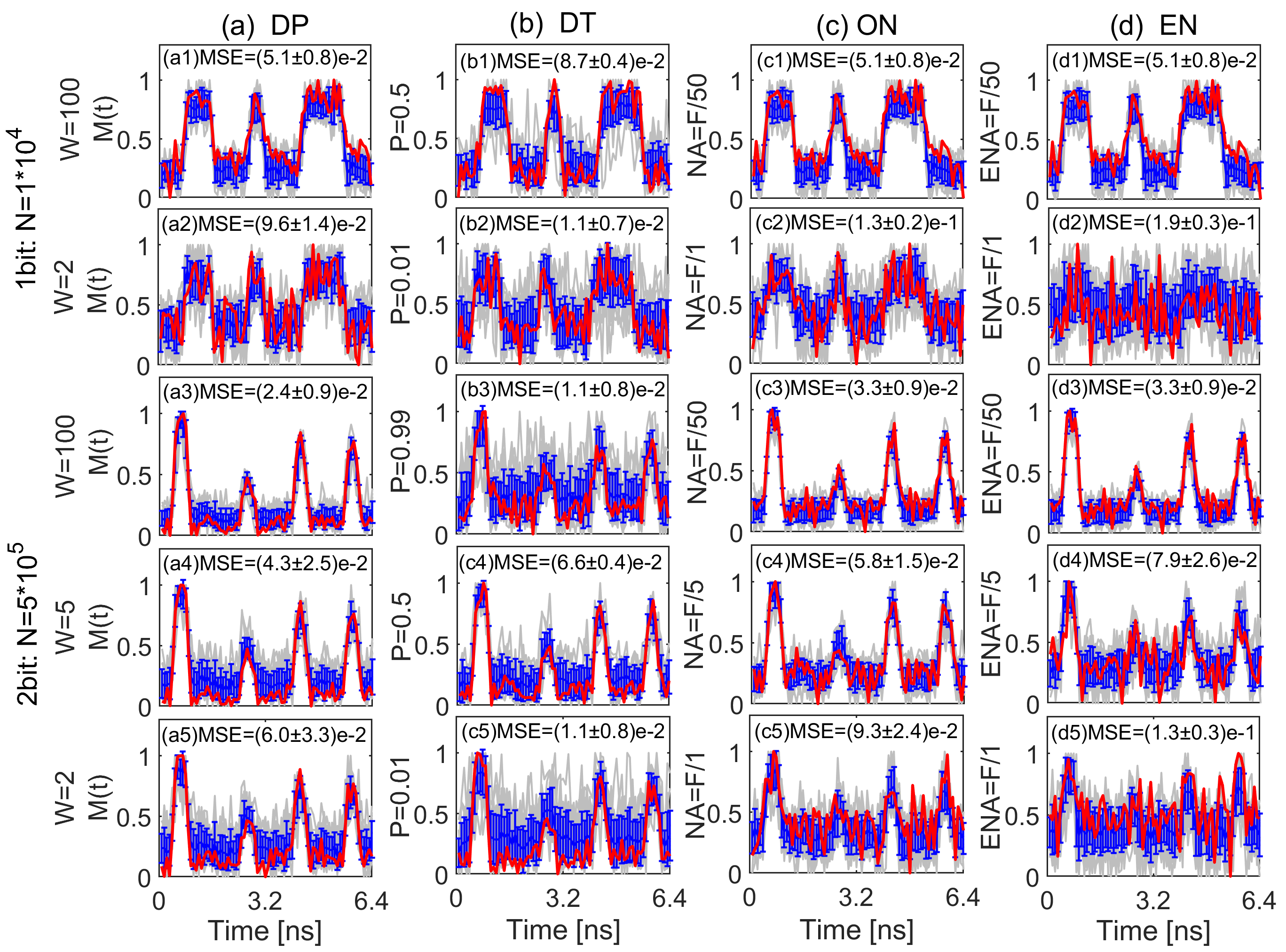}
	\caption{Image quality for detection precision (a), detection threshold (b), Gaussian noise (c) and electronic (d) for CTGI systems. $W$, detection precision; $P$, detection threshold; $NA$ and $ENA$, intensity amplitude of the noise, $F=max(I_{t})-min(I_{t})$; DP, detection precision; DT, detection threshold; ON, optical noise; EN, electrical noise; gray background, reconstructed signals of 20 measurements; blue line, mean values of the 20 measurements; red line, a typical signal acquisition. }
	\label{fig-3}
\end{figure*}

Intensity resolution is a critical factor influencing image quality, as shown in Fig. \ref{fig-3}(a). The detection accuracy $W$ decreases from high ($W=100$) to coarse ($W=2$). It can be observed that the MSEs are getting worse as the accuracy $W$ deteriorates. The MSE of the coarse detector is nearly twice that of the high-accuracy detector for both 1-bit and 2-bit signals. This experimental result substantiates the superiority of a high-accuracy detector in signal transmission, since it curtails data transmission time and augments system efficiency. When a coarse detector ($W=2$) is employed, which merely ascertains the presence or absence of light, the slow detector functions as a click detector, and the output signal of the detector is either 1 (arrive) or 0 (not arrive). This detection outcome is subject to the detection threshold. Values detected below the threshold are set to 0. Fig. \ref{fig-3}(b) illustrates the impact of the threshold $P$ of the click detector. The threshold $P$ denotes the normalized detection probability of the click detector. The optimal MSE is attained when the threshold line is set at $P=0.5$. If the threshold $P$ is extremely close to 0 or 1, it leads to a poor imaging effect. In this case, the temporal object can scarcely be reconstructed, thereby mandating an increased number of measurements and a prolonged decoding time to extract the information. The aforementioned phenomenon is especially prominent when transmitting 2-bit information. 

The robustness of decryption holds a pivotal position in dictating the precision and dependability of the ultimate information procured during the signal transmission process. Noise interference constitutes an inescapable factor that exerts an impact on the system. The information transmission system is prone to detection noise stemming from two sources: fluctuations in the light source and electronic noise arising from the photoelectric conversion of optical signals. Fig. \ref{fig-3}(c) depict the image quality upon the addition of Gaussian white noise to the ciphertext. The intensity resolution is set to high accuracy ($W=100$). It can be discerned that as the intensity of noise $NA$ escalates, a conspicuous deterioration in imaging quality occurs.  Optical noise exerts a substantial influence on statistical fluctuations, especially when the number of measurements $N$ is relatively scant. When the intensity amplitude of Gaussian noise approximates the fluctuation of signal's intensity $I_{t}$ ($NA=F$), the image quality experiences a severe decline. Furthermore, it can be noted that 2-bit transmission exhibits a more pronounced impact on statistical fluctuations. Fig. \ref{fig-3}(d) presents the decoding images with the incorporation of electrical noise. To facilitate precise comparisons regarding the noise effects, we persist in utilizing the intensity fluctuation of $I_{t}$ as a fundamental benchmark for analyzing electrical noise. It can be evidently observed that when the noise intensity approaches the signal intensity ($NA=F$), the MSE induced by electrical noise is substantially greater than that caused by optical noise. This observation intimates that the TGI transmission system manifests a more potent anti-interference proficiency against light noise.

\section{Multi-bit information transmission}
To conduct a more rigorous evaluation in our study, we employ the decoding accuracy rate (DAR) to quantify the quality of construction and assess the impact of decryption. DAR is calculated as 
\begin{equation}
	DAR=\sum {DM(t)_i}/ \sum {DM(t)_T}
	\label{EQ2}
\end{equation}
Where DM(t) is decryption signal, $\sum {DM(t)_i}$ represents the accurately decoded point, $\sum {DM(t)_T}$ is the total points of the decryption signal. It means that the higher the DAR, the higher the quality of decryption. 

\begin{figure*}[ht]
	\centering
	\includegraphics[width=0.98\textwidth]{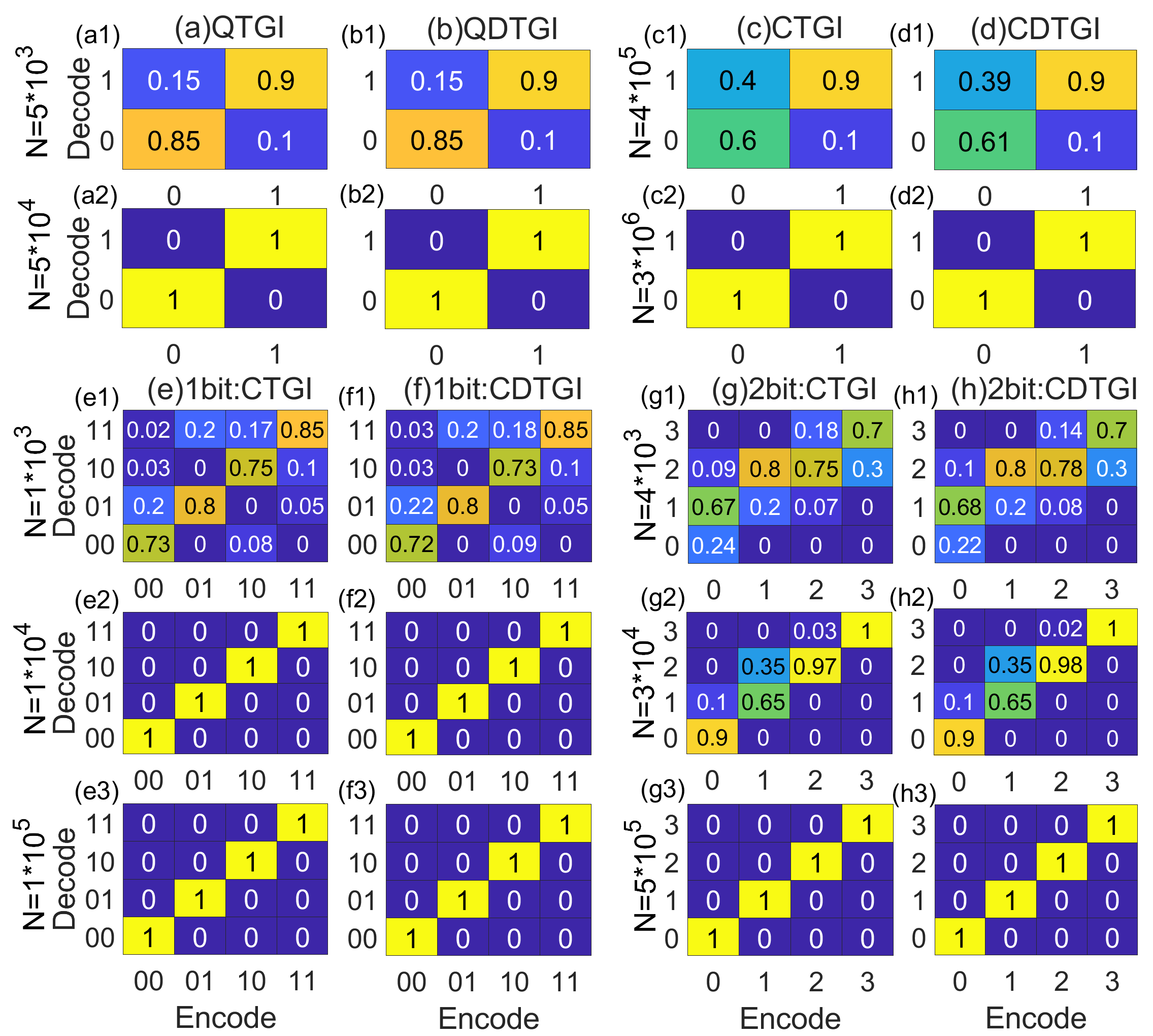}
	\caption{Decoding accuracy rate of the normalized 1-bit and 2-bit signals versus different measurements. Gray background, reconstructed signals of 20 measurements; blue line, mean values of the 20 measurements; red line, a typical signal acquisition.}
	\label{fig-4} 
\end{figure*}

The decoding results of the Q(D)TGI and C(D)TGI method is illustrated in Fig. \ref{fig-4}(a)-(d). It can be noted that the decoding quality demonstrates a remarkable enhancement as the number of measurements augments. When the reconstructed images align well with the target signals, the decoding accuracy rate (DAR) attains 100$\%$. The DARs of CDTGI are slightly better than those of CTGI when considering an equivalent quantity of data. However, the differential method does not possess an advantage in terms of decoding quality within the QTGI system. This is due to the fact that the post-selection detection probability of the fast detector (SNSPD) becomes 1 when the SNSPD acts as a click-or-not-click detector. Consequently, the image quality of the differential signal is equivalent to that of the QTGI. Both classical and quantum experimental demonstrate that differential signal acquisition does not exhibit a distinct advantage in the transmission system. It can be discerned from the first row in Fig. \ref{fig-2} that, despite the equivalence of the MSE values between QTGI and CTGI, QTGI presents markedly diminished fluctuation amplitudes. Consequently, the DAR of QTGI exceeds that of CTGI, as illustrated by the results in the first row of Fig. \ref{fig-4}, thereby providing empirical evidence of the enhanced performance characteristics of QTGI in the decoding process when compared to CTGI. Such a phenomenon might be attributed to the distinct signal processing mechanisms and error-correction capabilities inherent in QTGI. The lower fluctuation amplitudes in QTGI could potentially lead to a more stable decoding process, reducing the probability of bit errors and subsequently enhancing the overall DAR. Moreover, the utilization of quantum-related principles in QTGI, such as quantum entanglement or superposition, might confer additional advantages in terms of noise suppression and signal fidelity, which in turn contribute to the observed superiority in DAR over CTGI.

When engaged in the transmission of multi-bit information, as depicted in Fig. \ref{fig-4}, it can be evidently observed that the DAR of the C(D)TGI manifest a substantial augmentation as a greater number of measurements are procured. Moreover, the average DARs of the 2-bit signals are considerably lower than those of the 1-bit signals when the number of measurements $N<1\times10^5$. This disparity emanates from the significant background fluctuations that occur during the decoding process of the 2-bit signal, rendering it arduous to precisely discriminate between decoding 0 and 1. Hence, notwithstanding the approximately equivalent MSE (as illustrated in Fig. \ref{fig-2} (e)-(f)), a larger quantity of measurements is requisite to attain accurate decoding for the 2-bit signals.  Additionally, DTGI affords marginally enhanced decryption quality in comparison to TGI when furnished with an equivalent amount of data. However, as the measurement data becomes more copious, the discrepancy between the decoded images with and without differential acquisition dwindles. This is attributable to the fact that the differential acquisition bifurcates the signal into positive and negative constituents to obtain the differential signals, which attenuates the intensity fluctuations of the reconstructed images and augments the decoding accuracy rates relative to CTGI. However, when the reconstructed images with and without the differential method fit the target signals well, the decoding accuracy reach 100$\%$. Consequently, the advantage of the differential method becomes negligible.

\begin{figure*}[ht]
	
	\centering
	\includegraphics[width=0.98\linewidth]{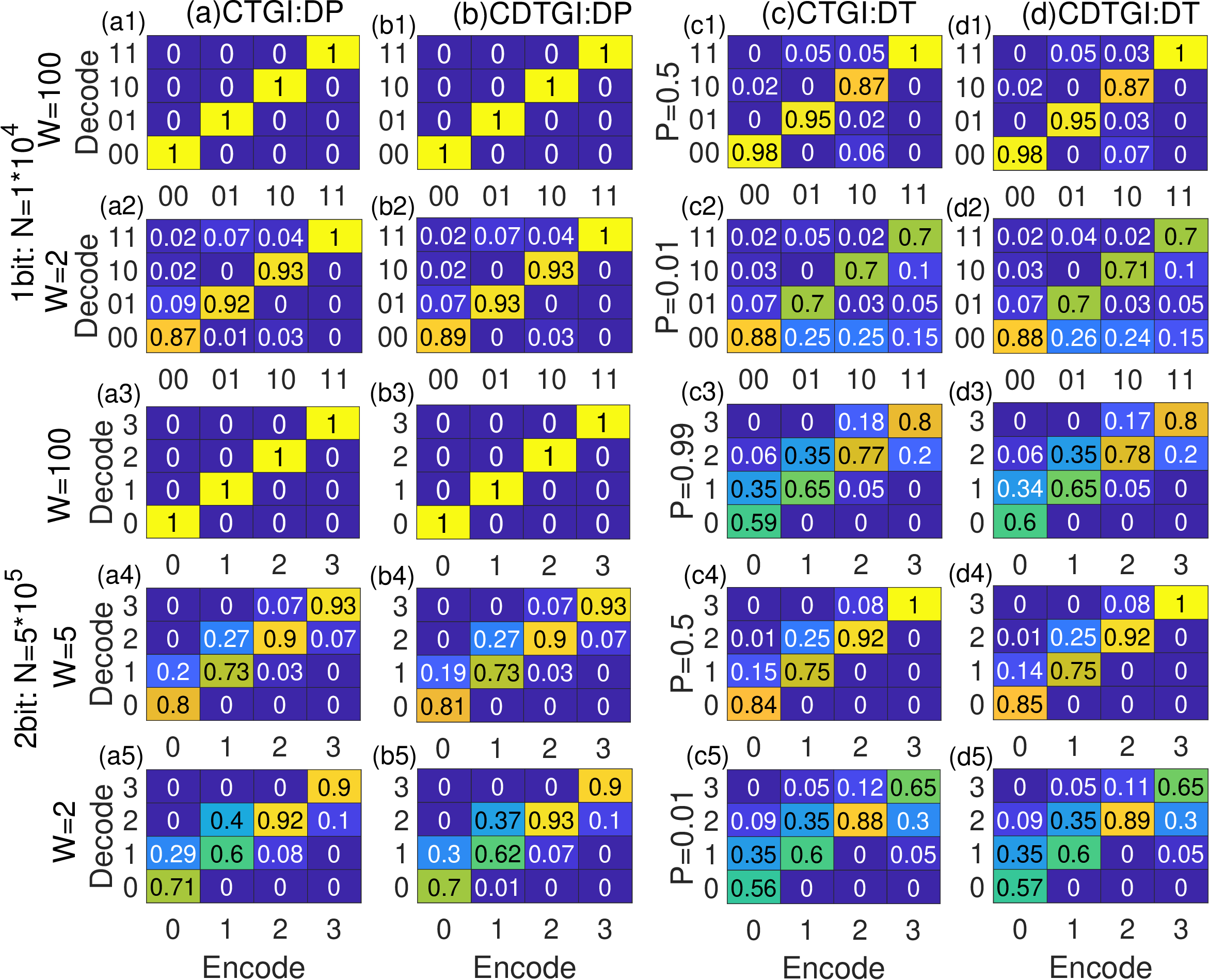}
	\caption{Decoding accuracy rate for CTGI systems with different detection accuracy and threshold. DP, detection precision; DT, detection threshold. }
	\label{fig-5}
\end{figure*}

The impact of the slow detector's detection accuracy $W$ on decoding quality is demonstrated in Fig. \ref{fig-5} (a) and (b). As $W$ decreases from high ($W=100$) to coarse ($W=2$), the DARs deteriorate. When the coarse detector ($W=2$) acts as a click detector, the average DAR is reduced by approximately $10 \%$ for the 1-bit signal and $20 \%$ for the 2-bit signal compared to that of the high-precision detector. The experimental results demonstrate that detectors with high accuracy can attain remarkable information decryption quality as the single-pulse information coding density increases. The fluctuation of the threshold $P$ for the coarse detector exerts a substantial impact on decoding quality,as depicted in Fig. \ref{fig-5} (c) and (d). Due to the detection results satisfying binary distribution, optimal decoding performance is achieved when threshold $P = 0.5$. If the threshold $P$ is in close proximity to 0 or 1, the average DAR is diminished by nearly $10 \%$ for the 1-bit signal and $20 \%$ for the 2-bit signal. This phenomenon can be ascribed to the fact that a reduction in the detection threshold heightens the detector's vulnerability to ambient noise, whereas an elevation in the detection threshold diminishes the detector's responsiveness to feeble signals. In such a scenario, the target signals are scarcely reconstructible, thereby mandating an increment in the number of measurements and a prolongation of the decoding time for information extraction. Consequently, the meticulous and strategic setting of the detection threshold during the transmission of multi-bit information emerges as a pivotal factor. It not only has the potential to markedly enhance decoding quality but also serves as a crucial safeguard against potential adversarial attacks that could otherwise imperil the integrity and fidelity of the transmitted data.

\begin{figure*}[ht]
	\centering
	\includegraphics[width=0.98\linewidth]{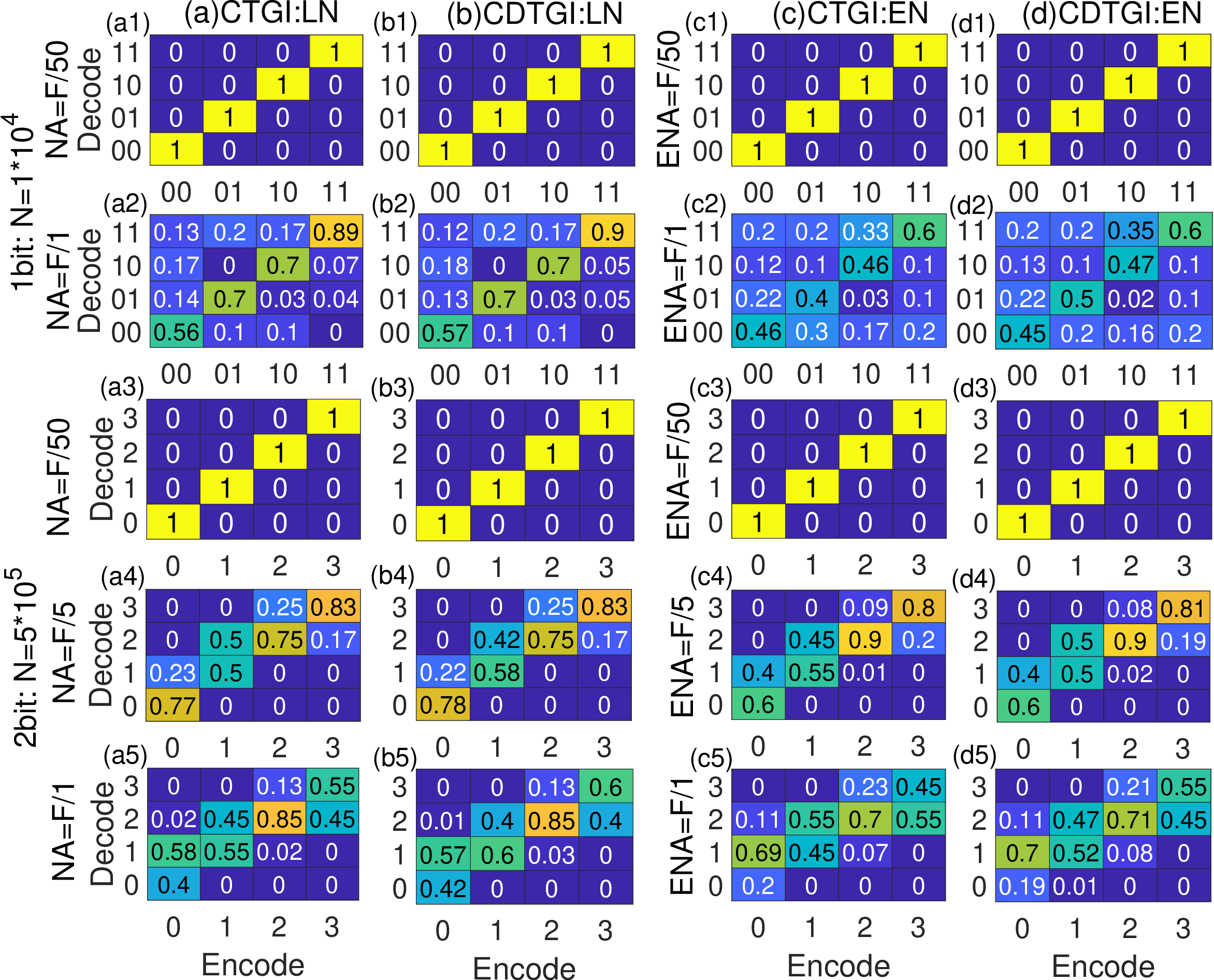}
	\caption{Decoding accuracy rate of Gaussian and electrical noise when the detection accuracy ($W=100$) is high. $NA$, intensity amplitude of the noise, $F=max(I_{t})-min(I_{t})$; ON, optical noise; EN, electrical noise.}
	\label{fig-6}
\end{figure*}

The decoding performance within the C(D)TGI system under the influence of light noise and electrical noise is depicted in Fig. \ref{fig-6}. The intensity resolution is set to high accuracy ($W=100$). It is apparent that with the increment of the noise intensity $NA$, the decoding performance deteriorates progressively. Specifically, when the amplitude of light noise reaches parity with the fluctuation of signal intensity $I_{t}$ ($NA=F$), the average DAR experiences a reduction of over 30$\%$ for 1-bit signals and in excess of 40$\%$ for 2-bit signals. Additionally, when the amplitudes of electrical noise are comparable to those of light noise, the average DAR declines by more than 50$\%$, which unambiguously indicates a relatively inferior performance in contrast to optical noise.  This result conclusively demonstrates that the C(D)TGI transmission system exhibits a more potent anti-interference capacity against light noise. Consequently, as the single-pulse information coding density augments, the influence of noise on transmission quality becomes more conspicuous. Although the differential method holds the potential to marginally enhance decoding quality, it fails to offer a substantial improvement. To achieve a more favorable decoding quality, supplementary measurements and extended sampling times are requisite. Moreover, noise may generate side-channel information, as documented in \cite{SC5,DeviceEvaluation}, which could potentially enable an adversary to intercept the bits without perturbing the dominant measurement results. Therefore, the possession of robust anti-noise capabilities is not a dispensable attribute but an essential prerequisite for ensuring the secure conveyance of information.

	\section{Conclusion}
In summary, this research make significant contributions to the field of optical information encryption through the implementation and analysis of QTGI and CTGI multi-information transmission systems. The experimental results demonstrate that the decoding accuracy rate exhibits a progressively enhanced sensitivity to the detection accuracy and detection threshold of the slow detector. In the context of transmitting multi-bit information, it is of utmost significance to institute an appropriate detection threshold and opt for a high-accuracy detector, thereby facilitating the attainment of superior decoding results. Furthermore, our information transmission system manifests robust anti-noise capabilities, effectively withstanding optical or electrical noise interference. Additionally, we have also demonstrated that the differential method has no optimization effect on the QTGI transmission system. However, a marginal enhancement in decoding quality is observed for the CTGI scheme. Our work has made significant contributions to the field of optical information encryption transmission by providing important theoretical guiding models. These advancements will help to enhance the security and reliability of optical communication systems and pave the way for future applications in areas such as optical communication and secure imaging. 
	\section*{Acknowledgments}
This work is supported by National Key Research and Development Program of China (Grant No. 2020YFA0309701), National Natural Science Foundation of China (Grant Nos. 62105004, 62371437).

\end{document}